\documentclass[%
 aip,
 amsmath,amssymb,
 reprint,%
]{revtex4-1}

\usepackage{graphicx}
\usepackage{dcolumn}
\usepackage{bm}

\usepackage[utf8]{inputenc}
\usepackage[T1]{fontenc}
\usepackage{mathptmx}
\usepackage{etoolbox}

\makeatletter
\def\@email#1#2{%
 \endgroup
 \patchcmd{\titleblock@produce}
  {\frontmatter@RRAPformat}
  {\frontmatter@RRAPformat{\produce@RRAP{*#1\href{mailto:#2}{#2}}}\frontmatter@RRAPformat}
  {}{}
}%
\makeatother
\begin{document}

\preprint{AIP/123-QED}

\title{Collective oscillations in the finite-size Kuramoto model below the critical coupling: shot-noise approach}
\author{S. Yu. Kirillov}
 \email{skirillov@ipfran.ru}
 \affiliation{A.V. Gaponov-Grekhov Institute of Applied Physics of the Russian Academy of Sciences, Ulyanova Street 46, Nizhny Novgorod 603950, Russia
}%
\author{V. V. Klinshov}%
 \email{vklinshov@hse.ru}
  \affiliation{A.V. Gaponov-Grekhov Institute of Applied Physics of the Russian Academy of Sciences, Ulyanova Street 46, Nizhny Novgorod 603950, Russia
}%
 \affiliation{National Research University Higher School of Economics, 25/12 Bol’shaya Pecherskaya Street, Nizhny Novgorod 603155, Russia
}%



\begin{abstract}
The Kuramoto model, a paradigmatic framework for studying synchronization, exhibits a transition to collective oscillations only above a critical coupling strength in the thermodynamic limit. However, real-world systems are finite, and their dynamics can deviate significantly from mean-field predictions. Here, we investigate finite-size effects in the Kuramoto model below the critical coupling, where the infinite-size theory predicts complete asynchrony. Using a shot-noise approach, we derive analytically the power spectrum of emergent collective oscillations and demonstrate their dependence on the coupling strength. Numerical simulations confirm our theoretical results, though deviations arise near the critical coupling due to nonlinear effects. Our findings reveal how finite-size fluctuations sustain synchronization in regimes where classical mean-field theories fail, offering insights for applications in neural networks, power grids, and other coupled oscillator systems. 
\end{abstract}

\maketitle

\begin{quotation}
The Kuramoto model is a classical object of nonlinear dynamics, widely used to study synchronization in systems of coupled oscillators -- from neurons firing in the brain to generators in power grids. It captures how individual units with different natural frequencies can spontaneously align their rhythms through interactions, leading to collective oscillations when the coupling exceeds a critical value. While there is a well-established theory of Kuramoto model for infinitely large (thermodynamic limit) systems, real-world networks are finite, and their dynamics often deviate from mean-field approximations. The present paper investigates how finite-size effects enable synchronization even in regimes where the thermodynamic theory predicts the asynchronous state. We develop a theory based on the so-called shot-noise approach which allows to accurately calculate the power spectrum and the magnitude of collective oscillations below the critical coupling. Our findings advance the understanding of synchronization in real systems, where finite-size effects may play pivotal role.
\end{quotation}

\section{Introduction}

Complex networks consisting of many interacting active units are omnipresent in nature and technology. Examples include neuronal structures in the brain \cite{Wilson2022,Breakspear2010}, arrays of coupled Josephson junctions \cite{Wiesenfeld2020}, telecommunication networks \cite{Nguefoue2023}, power grids \cite{Filatrella2008,Khramenkov2025}, and many others \cite{Mehta2010,Kasatkin2021,Munyaev}. Notably, Ljup\v{c}o Kocarev made significant contributions to the understanding of synchronization and control in complex networks, providing valuable insights into their dynamical behavior \cite{Kocarev,Trpevski,Mishkovski,Tang}.

Research on complex dynamical networks primarily focuses on identifying collective (macroscopic) activity patterns that characterize their global behavior without delving into the detailed (microscopic) dynamics of individual elements. Contrary to the intuitive expectation that adding more elements increases system complexity, a sufficient increase in size may instead lead to a simplification of collective dynamics. In such cases, the behavior can be described using just a few averaged variables governed by the so-called mean-field models \cite{Carlu,Volo,Byrne,Spaeth,Bick2020}. For homogeneous populations, the dynamics can be explicitly reduced to low-dimensional models even for a finite number of units \cite{Watanabe1994, Pikovsky2008, Kirillov2020}. For heterogeneous populations, analytical treatment is possible in the thermodynamic limit of an infinite number of units. If the heterogeneity arises from additive internal noise in individual elements, the system's dynamics can first be reduced to a Fokker-Planck equation, whose approximate solutions can, in turn, be described by a low-dimensional system of ordinary differential equations for the Kuramoto-Daido order parameters \cite{Sakaguchi1988, Daido1996, Klinshov2021}. For the heterogeneity in local parameters of the units, reduction is possible using the Ott-Antonsen ansatz \cite{Ott2008, Ott2009} and the related Lorentzian ansatz \cite{Montbrio2015}. These reduction methods have proven highly effective in studying dynamics of neural populations \cite{Coombes2023, Pyragas2021, Nandi2024, Gast2024} and large-scale neural networks \cite{Gerster,Badarin}, including the effects of interaction delays \cite{Lee2009, Laing2016, Wolfrum2022, Devalle2018} and adaptive coupling \cite{Duchet2023, Fennelly2025, Eydam2024, Pietras2025}.  In cases where the both local parameter heterogeneity and noise sources are present, the reduction can be performed using the circular cumulant method \cite{Tyulkina2018, Goldobin2021, Ageeva2025,Pyragas2024}. 

Mean-field models have driven tremendous progress in studying  dynamics of large-scale populations. However, it is important to emphasize that their derivation relies on the infinite-size approximation, whereas real populations have finite size. This raises further questions about finite-size effects and their role in collective dynamics \cite{Schwalger,Goldobin2024, Ageeva2024, Goldobin2025, Zhang2025}.  
Recently, we introduced an approach to studying finite-size effects based on the  concept of ``shot noise'' \cite{Klinshov2022, Klinshov2023, Kirillov2023, Kirillov2024}, a term introduced by analogy with shot noise in electrical circuits. According to this concept, the finite-size population shows fluctuations near the solution of the infinite-size population, which can be described using stochastic mean-field models. This approach allowed to derive the power spectrum of the shot noise and explain its effect on the collective dynamics of the finite-size populations. In particular, it allowed to predict large-amplitude collective oscillations and recurrent switching between metastable states in neural networks -- the phenomena not captured by the classical mean-field models based on the thermodynamic limit. It was shown that the shot noise is strongly colored and can exhibit prominent peaks which underlines the importance of the study of its spectrum.

Previously, we applied our shot-noise approach only to networks of quadratic integrate-and-fire neurons. In this work we extend it to a different system, namely, to the classical Kuramoto model describing weakly coupled phase oscillators \cite{Kuramoto1984,Rodrigues,Lu,Acebron}. In the thermodynamic limit, this system demonstrates a soft transition from complete asynchrony to partial synchrony manifesting itself as collective oscillations. We demonstrate that in a finite-size system, collective oscillations exist even for the subcritical coupling. We study the properties of these oscillations in the subcritical region and derive analytically their power spectrum depending on the coupling strength.   

\section{Model}
Consider a population of $N$ Kuramoto oscillators
\begin{equation}\label{a1_01}
\frac{d\theta_j}{dt} = \omega_j + \frac{1}{N}\sum_{j=1}^N K_{j,k} \sin(\theta_k - \theta_j), \quad j = 1, \dots, N,
\end{equation}
where  $\theta_j\in [0, 2\pi)$ is the phase of the $j$-th oscillator, $\omega_i$ is its natural frequency,  $K_{j,k}$ is the coupling strength from the $k$-th oscillator to the $j$-th oscillator.
Now, assume that the coupling strength is uniform across all pairs of oscillators: $K_{j,k} = K$. Then it is convenient to use the Kuramoto order parameter 
\begin{equation}\label{a1_03}
R = \frac{1}{N} \sum_{j=1}^N e^{i \theta_j},
\end{equation}
and the system dynamics can be rewritten as 
\begin{equation}\label{a1_01}
\frac{d\theta_j}{dt} = \omega_j + K \operatorname{Im}  R e^{-i\theta_i}.
\end{equation}

\section{Thermodynamic limit}

First let us consider the system in the thermodynamic limit $N\rightarrow \infty$, then it can be described by continuous density of  $f(\omega,\theta,t)$ satisfying the continuity equation
\begin{equation}\label{a2_03}
\frac{\partial f}{\partial t}+\frac{\partial}{\partial\theta}\bigg\{\Big[\omega+\frac{K}{2i}(r e^{-i\theta}-r^* e^{i\theta})\Big]f\bigg\}=0,
\end{equation}
where $r\equiv R$, but here and further we will use the lowercase $r$ for the \textit{infinite-size} population whose the order parameter is given by 
\begin{equation}\label{a2_04}
r=\int_0^{2\pi}d\theta\int_{-\infty}^\infty d\omega f (\omega,\theta,t) e^{i\theta}.
\end{equation}

Now let us write the density function in the form of a Fourier series
\begin{equation}\label{a2_05}
f=\frac{g(\omega)}{2\pi}\bigg\{1+\sum_{n=1}^\infty \Big[ f_n(\omega,t)e^{i n\theta}+c.c.\Big]\bigg\},
\end{equation}
where $c.c.$ denotes complex conjugate. According to Ott-Antonsen anzats, the solution of \eqref{a1_03} can be found in the class
\begin{equation}\label{a2_06}
f_n(\omega, t)=\big[\alpha(\omega, t)\big]^n,
\end{equation}
where the $\alpha(\omega, t)$ evolves according to
\begin{equation}\label{a2_07}
\frac{\partial\alpha}{\partial t}+\frac{K}{2}\big(r\alpha^2-r^*\big)+i\omega\alpha=0,
\end{equation}
and the order parameter is defined as
\begin{equation}\label{a2_08}
r=\int_{-\infty}^{\infty}d\omega\alpha^*(\omega,t)g(\omega),
\end{equation}
where 
\begin{equation}\label{a2_01}
g(\omega) = \int_0^{2\pi}f(\omega,\theta,t)d\theta
\end{equation}
is the (stationary) distribution of the frequencies. 

Now let us assume that this distribution is Lorentzian
\begin{equation}\label{a2_09}
g(\omega)=\frac{1}{\pi}\frac{\Delta}{(\omega-\omega_0)^2+\Delta^2}.
\end{equation}
Without the loss of generality we can set $\omega_0=0$ and $\Delta=1$, then  the integral  (\ref{a2_08}) can be evaluated via the residue theorem as $r=\alpha^*(-i,t)$. Substituting  $\omega=-i$ into (\ref{a2_07}) one obtains a closed mean-field system for the Kuramoto order parameter
\begin{equation}\label{a2_10}
\frac{dr}{dt}=\bigg(\frac{K}{2}-1\bigg)r-\frac{K}{2}r^3.
\end{equation}

This equation has a stable steady state $r=0$ for weak coupling $K<2$ which destabilizes and gives rise to a stable limit cycle for $K>2$ which corresponds to the Kuramoto transition to synchronization at the critical coupling strength $K=2$. Further we will consider only subcritical case with $K<2$ so that the population is asynchronous and its order parameter equals zero. Note however that this is true only for the thermodynamic limit $N\to\infty$, while the finite-size population has non-zero order parameter even for weak coupling. To show this, let us start from the simplest case of zero coupling $K=0$.

\section{Uncoupled finite-size population}

Let us consider the case where the number of network elements is large but finite  $1\ll N < \infty$, and  the natural frequencies of the oscillators $\omega_i$ are distributed randomly within a given distribution form (\ref{a2_09}). In the absence of coupling ($K=0$) the individual phases rotate uniformly, so that $\theta_j(t)=\omega_j t +\theta_{j0}$. Then the order parameter can be expressed as 
\begin{equation}\label{a3_02}
s(t)=\frac{1}{N}\sum_{j=1}^N e^{i\omega_j t+i\theta_{0j}},
\end{equation}
where $s(t)\equiv R(t)$, and here and further we will use $s(t)$ to denote the order parameter of the \emph{finite-size} population. Obviously, $s(t)\neq 0$, and for incommensurable natural frequencies it shows quasi-periodic oscillations, which resembles a stochastic signal  for large number of units. We in interpret this signal as a shot noise $\chi(t)$ added to the signal of the infinite-size population
\begin{equation}\label{eq:shotnoisedef}
    s(t)=r(t)+\chi(t).
\end{equation}
In the case of zero coupling, $r(t)\equiv 0$, and $\chi(t)\equiv \chi_0(t)$, where the subscript ``0'' refers to the absence of coupling in which case we call the shot noise ``free''.

Let us calculate the power spectrum of the free shot noise. Its autocorrelation function equals
\begin{multline}\label{a3_03}
K_0(\tau)=\langle\ R(t)R^*(t-\tau)\rangle=\\
= \lim_{T\rightarrow\infty}\frac{1}{N^2 T}\int_0^T\sum_{j,k=1}^N e^{i(\omega_j-\omega_k)t+i(\theta_{0j}-\theta_{0k})+i\omega_k\tau}dt.
\end{multline}

\begin{figure*}[t]
\center{\includegraphics[width=\textwidth]{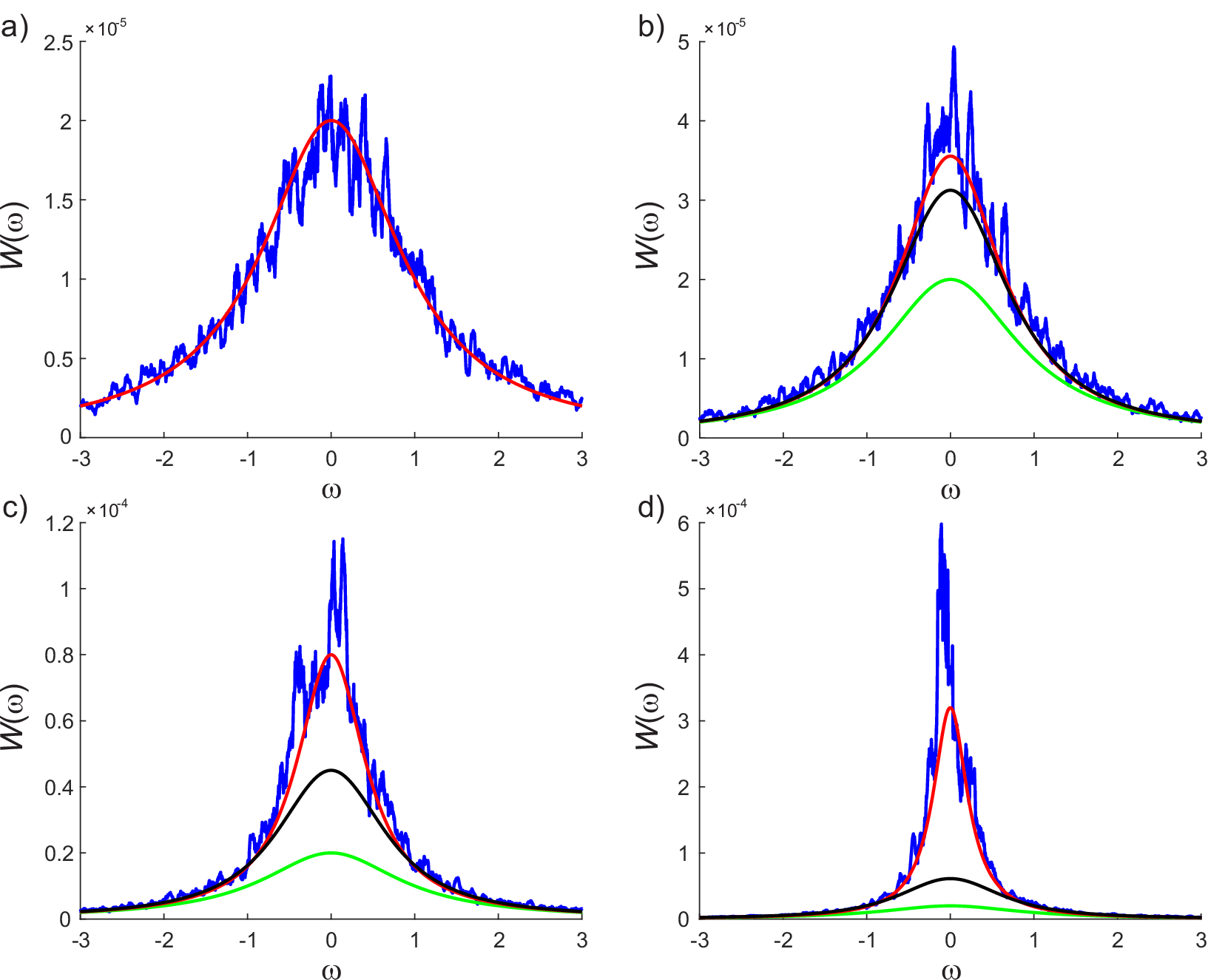}}
\caption{Power spectra  of collective oscillations of a system Kuramoto system with $N=10^5$ units for the coupling strength (a) $K=0$, (b) $K=0.5$, (c) $K=1$ and (d) $K=1.5$. Blue lines: results of the numerical simulations. Green lines: power spectra of the free shot noise given by \eqref{a3_05}. Black lines: power spectra given by Eq. \eqref{a4_16}.  Red lines: power spectra given by Eq. \eqref{a4_16_1}. In panel (a), the green, the black and the red lines coincide since $K=0$. Mind the difference in the scale of the vertical (but not horizontal) axis between the panels.} \label{fig:1}
\end{figure*}

It is easy to see that in sum (\ref{a3_03})  the contribution is nonzero only for  terms with $j=k$ so that
\begin{equation}\label{a3_04}
K_0(\tau)=\lim_{T\rightarrow\infty}\frac{1}{N^2 T}\int_0^T\sum_{j=1}^N e^{i\omega_j\tau}dt=\frac{1}{N^2}\sum_{j=1}^N e^{i\omega_j\tau},
\end{equation}
and the power spectrum of the free shot noise
\begin{equation}\label{a3_05} 
W_0(\omega)=\int_{-\infty}^{\infty}K(\tau)e^{-i\omega\tau}d\tau=\frac{2\pi}{N^2}\sum_{j=1}^N\delta(\omega-\omega_j)\approx\frac{2\pi}{N}g(\omega). 
\end{equation}
This power spectrum is illustrated in Fig. 1(a) together with the results of the numerical simulations. For the simulations we took $N=10^5$ units with natural frequencies randomly drawn from the distribution \eqref{a2_09} and initial phases randomly drawn from the uniform distribution $\theta\in[0;2\pi)$. We integrated the population for $t=10^5$ using the Euler method with the time step $\Delta t=10^{-5}$, then calculated the power spectrum using FFT and applied moving average smoothing with a window frequency window  $\Delta \omega=2\pi 10^{-2}$. Numerical results are in good agreement with the theoretical prediction.

\section{Finite-size population with subcritical coupling}

Let us now study how the shot noise changes when the coupling in the population becomes nonzero although subcritical ($0<K<2$). Note that in this case the infinite-size population would be  still in the trivial steady state($r=0$). However, the dynamics of the finite-size population change due to the nonzero coupling since the free shot noise is fed back to the population which modifies its power spectrum. The analytical treatment of this problem can be performed using the so-called ``nested'' configuration suggested by us in \cite{Klinshov2022}. In this setting a population of finite size $N$ is represented as part of a population network $N^+\to\infty$ containing an infinite number of elements (see Fig. 2 for illustration). It is assumed that the  output of the small population affects the entire large population (including the small population). Such consideration preserves the dynamics of the small population, while the large population can be accurately described using the Ott-Antonsen approach. Indeed,  the dynamics of the infinite-size population is given by
\begin{equation}\label{a1_01_1}
\frac{d\theta_j}{dt} = \omega_j + K \operatorname{Im} s(t) e^{-i\theta_j},
\end{equation}
where $s(t)$ is the output (order parameter) of the finite-size population. Using the same technique as in Sec. 3, this dynamics can be reduced to the following mean-field system for the order parameter $r(t)$:
\begin{equation}\label{a2_10_1}
\frac{dr}{dt}=-r+\frac{K}{2}(sr^2-s^*).
\end{equation}

\begin{figure}[t]
\includegraphics[width=7cm]{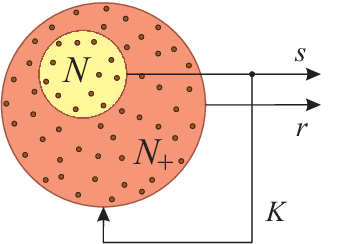}
\caption{The ``nested'' configuration: the population of the size $N$ is considered as a part of the larger population of the size $N_+\to\infty$. The signal $s$ from the smaller population acts on the larger population (including the smaller one inside it).} \label{fig:2}
\end{figure}

For large $N$ the shot noise is weak $\sim N^{-1/2}$ which allows to linearize \eqref{a2_10_1} near $r=0$ and calculate the linear frequency response
\begin{equation}\label{a2_10_2}
S(\omega)=\frac{1}{1+i\omega}.
\end{equation}
Following the approach developed in \cite{Klinshov2022} for neural mass models, we can calculate the Fourier spectrum of the fluctuations, induced by the free shot noise from the finite population in the infinite population as $S(\omega)w_0(\omega)$, where $w_0(\omega)$ is the Fourier spectrum of the free shot noise. Then, using the definition of the shot noise  \eqref{eq:shotnoisedef} we obtain the following power spectrum of the full shot noise
\begin{equation}\label{a4_16}
W(\omega)=\bigg|1+\frac{K}{2}S(\omega)\bigg|^2 W_0(\omega)=\frac{2\pi}{N}\bigg|1+\frac{K/2}{1+i\omega}\bigg|^2 g(\omega).
\end{equation}

However, the former approach assumes that the shot noise is much smaller than the signal from the infinite population which is true for neural masses but false for the subcritical Kuramoto system whose output is close to zero. Thus, rather than considering the fluctuations induced by the free shot noise one should write the following self-consistency condition:
\begin{equation}\label{a2_10_2}
w(\omega)=\frac{K}{2}S(\omega)w(\omega)+w_0(\omega),
\end{equation}
Solving this equation one finally arrives to the following expression for the power spectrum of the shot noise at the subcritical coupling strength $K$:
\begin{equation}\label{a4_16_1}
W(\omega)=\bigg|\frac{1}{1-\frac{K}{2}S(\omega)}\bigg|^2 W_0(\omega)=\frac{2\pi}{N}\bigg|\frac{1+i\omega}{1+i\omega-K/2}\bigg|^2 g(\omega).
\end{equation}

Panels (b), (c), (d) in Fig. 1 show the power spectra of the shot noise for increasing values of the coupling strength $K=0.5$, $1$, $1.5$. The numerical results were obtained in the same way as for the system with $K=0$. The agreement between the numerically obtained and theoretically predicted power spectra is remarkable although deteriorates with the growth of the coupling strength. Namely, for $K=1$ and especially for $K=1.5$ narrow high peaks appear in the numerical spectrum although the theoretical one remains smooth and uni-modal. The reason is that the mean-field system \eqref{a2_10} approaches Andronov-Hopf bifurcation at which the linearization becomes inadequate. Note also that the Eq. \eqref{a4_16_1} diverges at $K=2$ (for $\omega=0$) which further indicates the inapplicability of our approach in the vicinity of the Kuramoto transition.

\begin{figure}
\center{\includegraphics[width=8cm]{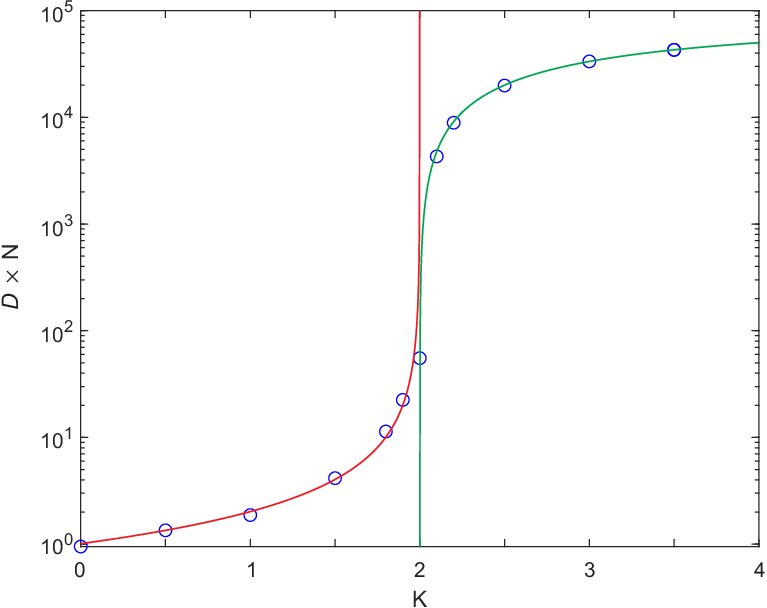}}
\caption{The variance of the Kuramoto order parameter $D$ versus the coupling strength $K$ for $N=10^5$. Blue circles: results of the numerical simulations. Red line: the value given by Eq. \eqref{a4_18_1} for $0<K<2$. Green line: the value given by Eq. \eqref{a4_19} for $K>2$. Note the logarithmic scale of the vertical axis. Note also that the variance $D$ is multiplied by $N$.} \label{fig:Dispersion}
\end{figure}

In order to better understand how the accuracy of our theory changes with the increase of coupling, we studied how the magnitude of collective oscillations depends on the coupling strength. For this sake we calculated the variance of the order parameter numerically for different $K$ and compared it to the theoretical value which can be estimated from the Parseval's theorem as
\begin{equation}\label{a4_18}
D=\frac{1}{2\pi}\int_{\infty}^{\infty}W(\omega)d\omega.
\end{equation}
Using Eq. \eqref{a4_16} for the power spectrum of the shot noise, this integral for $K<2$ becomes:
\begin{equation}\label{a4_18_1}
D=\frac{1}{N}\frac{1}{1-K/2}.
\end{equation}
Remarkably, our results are in perfect agreement with formula (3.12) from \cite{Daido}, despite being obtained through an entirely different approach.

The theoretic value of the variance $D$ is plotted versus the coupling strength $K$ in Fig. 3 in the logarithmic scale. For subcritical coupling we use $0<K<2$ we use Eq. \eqref{a4_18_1}, while for supercritical coupling $K>2$ we use the prediction of the thermodynamic mean-field model \eqref{a2_10} which gives
\begin{equation}\label{a4_19}
    D=1-2/K.
\end{equation}
The numerical results are in perfect agreement with our theory for $K$ as large as $1.9$. Our theory breaks only very close to the critical coupling  $K=2$ where formulae \eqref{a4_18} diverges. Recall that the thermodynamic theory predicts that the collective oscillations are strictly zero for $K\leq 2$, while our approach allows to expand the theoretical description of collective oscillations to subcritical couplings.

\section{Conclusions and discussion}

The present study demonstrates that finite-size populations of Kuramoto oscillators exhibit collective oscillations even below the critical coupling strength ($K<2$), where the thermodynamic limit predicts complete asynchrony $( r = 0 )$. These oscillations arise due to ''shot noise``,  stochastic fluctuations inherent to finite systems.  We derive an analytical expression for the power spectrum of these oscillations, showing how it depends on the coupling strength. Numerical simulations confirm the theoretical predictions, though deviations increase as the coupling approaches the critical value due to the breakdown of linear approximations.  Namely, the theory predicts a divergence in the power spectrum at the critical coupling $K = 2$ highlighting the limitations of the approach near the Kuramoto transition point, where nonlinear effects dominate.  

Finite-size effects in Kuramoto model have been explored in a number of prior works \cite{Hong,Coletta,Lee,Park}. However, to the best of our knowledge we are the first to calculate the power spectrum of finite-size induced fluctuations. Our results show that the shot-noise framework, previously applied to neural networks, can be  successfully extended to other system, which underscores its generality. The divergence close to the critical transition point aligns with the expected breakdown of linearized models near bifurcations, emphasizing the need for nonlinear analyses in future work. Another intriguing challenge is to extend the analysis to the super-threshold case when the mean-field model shows finite-amplitude periodic oscillations. 

To conclude, the present work extends the shot-noise approach, previously used only for  networks of quadratic integrate-and-fire neurons, to different systems. It  bridges the gap between infinite-size mean-field theories and real-world finite systems, showing that finite-size fluctuations can induce collective dynamics absent in thermodynamic limit predictions, and opens  avenues for understanding finite-size effects in diverse oscillator systems, with implications for neuroscience, engineering, and physics. 

\begin{acknowledgments}
The research was carried out within the framework of the state assignment of the A.V. Gaponov-Grekhov Institute of Applied Physics of the Russian Academy of Sciences, scientific topic FFUF-2024-0011.
\end{acknowledgments}

\section*{Data Availability Statement}

The data that support the findings of this study are available from the corresponding author upon reasonable request.


\section*{references}

\end{document}